\documentstyle{article}
\title{
\begin{flushright}
{\bf\normalsize   COLO-HEP-328 \\ LPTHE-Orsay-93/44 }  \\
\end{flushright}
\vskip 10pt
\bf Scaling in Steiner Random Surfaces
}

\author{ {\it C.F. Baillie} \\
         Dept of Computer Science, University of Colorado\\
         Boulder, CO 80309, USA\\ \\
         {\it A. Irb\"ack}\\
         Department of Theoretical Physics\\
         University of Lund\\
         S\"olvegatan 14A, S-223 62 Lund, Sweden\\ \\
         {\it W. Janke}\\ Institut fur Physik\\ Johannes
         Gutenberg Universitat\\
         D 55099 Mainz, Germany\\ \\
         {\it D.A. Johnston}\\
         LPTHE\\
         Universite Paris Sud, Batiment 211\\
         F-91405 Orsay, France$^{1}$\\
}

\begin{document}
  \maketitle
                      {\Large
                      \begin{abstract}
%
It has been suggested that the modified Steiner action functional has desirable
properties for a random surface action.
In this paper we investigate the
scaling of the string tension and massgap in
a variant of this action on
dynamically triangulated random surfaces
and compare the results with
the gaussian
plus extrinsic curvature actions that have been used previously.
\\ \\
Submitted to Phys Lett B.   \\ \\
$~^{1}$ {\it Address:} Sept. 1993 - 1994, {\it Permanent Address:} Maths Dept,
Heriot-Watt
University, Edinburgh, Scotland \\
%
                        \end{abstract} }
%
  \thispagestyle{empty}
%
%
  \newpage
%
                  \pagenumbering{arabic}

\section{Introduction}
The issue of whether a non-trivial
continuum limit exists for
gaussian plus extrinsic curvature (GPEC) lattice actions
of the form
\begin{equation}
S = \sum_{<ij>} ( \vec X_i - \vec X_j )^2 +
\lambda \sum_{\Delta_i, \Delta_j} ( 1 - \vec n_i  \cdot \vec n_j )
\label{e01}
\end{equation}
on dynamically triangulated random surfaces,
is of interest for the construction of well-defined lattice
versions of string theory \cite{1,2,2a,3,4} as well as for
constructing models of membranes in biophysics and chemistry.
The second term in equ.(1),
where the $\vec n_i$ are the unit normals on neighbouring
triangles, is a discretization of the extrinsic
curvature and acts as a ``stiffness'' term.
If this term is absent one has
a gaussian discretization of the basic Polyakov action
\cite{4a} which gives rise to pathologically crumpled surfaces
due to the failure of the string tension to scale \cite{4b}.
The dynamical nature of the triangulation is manifested as a
sum over triangulations, $\sum_{T}$, in the
canonical (fixed number of points) partition function
\begin{equation}
Z_N(\lambda) = \sum_{T} \int \prod_{i=1}^N d \vec X_i \delta ( \sum_i \vec X_i)
\exp (- S),
\label{e02}
\end{equation}
where the delta function is inserted to kill
the translational zero-mode,
and $N$ is the number of points.
This means that we have in effect a fluid
surface.
The GPEC model of equs.(1,2) apparently has a low $\lambda$ crumpled phase
and a large $\lambda$ smooth phase similar to those displayed
by identical models on fixed triangulation surfaces \cite{5}
where the sum over triangulations in equ.(2) is dropped.
The initial work in \cite{1} found a pseudo-second order
transition on small lattices, but later work \cite{2,3,4} with larger
lattices and better statistics suggested rather that the transition
was higher order, or a crossover phenomenon \cite{4,6}.

The strongest evidence so far that there {\it is}, indeed, a transition
comes from the measurements of the scaling of the string tension
and mass gap carried out in \cite{2}.
An earlier measurement of the string tension also found results
that were consistent with scaling, but in this the
points on the boundary, which constituted a
large proportion of the total number of points, were
physically pinned down \cite{2a} and the vanishing of the
lattice string tension at the critical point was assumed.
Although
analytical calculations suggested \cite{9} that the extrinsic curvature
coupling $\lambda$ in equ.(1) was asymptotically free
and hence that there was no non-trivial theory for finite
$\lambda$ in the lattice action, the measurements in \cite{2}
were strongly indicative
of scaling and hence a finite string tension in physical units.
This implies a non-trivial continuum limit at finite $\lambda$.

Pending clarification of the behaviour of the GPEC action on
dynamical triangulations, we thought it a worthwhile exercise
to investigate possible alternative lattice random surface actions
in order to see if their behaviour was more (or less!) clear-cut.
We have already conducted some preliminary simulations \cite{10,10a}
of actions containing terms of the form suggested by
Savvidy et.al. \cite{11,11a} that incorporate the
modified Steiner functional \cite{12}.
The basic ``Steiner'' action is just
\begin{equation}
S_{Steiner} = {1 \over 2} \sum_{<ij>} | \vec X_i - \vec X_j | \theta
(\alpha_{ij}),
\label{e4a}
\end{equation}
where
$\theta(\alpha_{ij}) = | \pi - \alpha_{ij} |$
and $\alpha_{ij}$ is the angle between the
embedded neighbouring triangles with common link $<ij>$.
This is essentially a coarse discretization of the absolute value of the
trace of the second fundamental form
of the surface, rather than its square which appears in the GPEC action.
It was observed in
\cite{13} that an action containing only
this term ran into problems with the entropy
of vertices in
smooth configurations and failed to give a well-defined grand canonical
(varying number of vertices)
partition function. It is a relatively simple matter however to concoct
variations on this theme that constrain the errant planar vertices
somewhat such as
\begin{equation}
S_1 = {1 \over 2} \sum_{<ij>} | \vec X_i - \vec X_j | + { \lambda \over 2}
\sum_{<ij>} | \vec X_i - \vec X_j |
\theta (\alpha_{ij})
\label{e5}
\end{equation}
or even
\begin{equation}
S_2 = \sum_{\Delta} |\Delta | + { \lambda \over 2} \sum_{<ij>} | \vec X_i -
\vec
X_j | \theta (\alpha_{ij}),
\label{e6}
\end{equation}
where the $| \Delta |$ is just the area of triangle $\Delta$
as seen in the space in which the surface is embedded. In \cite{11a}
another alternative was suggested in which $\theta$ was modified
to $\theta(\alpha_{ij}) = | \pi - \alpha_{ij} |^\xi$ with $\xi<1$
which also appeared to improve the convergence of the grand-canonical
partition function.

In \cite{10} we carried out simulations of $S_1, S_2$
along with a further permutation
combining a gaussian term with the Steiner part
\begin{equation}
S_3 = {1 \over 2} \sum_{<ij>} ( \vec X_i - \vec X_j )^2 + { \lambda \over 2}
\sum_{<ij>} | \vec X_i - \vec X_j |
\theta (\alpha_{ij})
\label{e7}
\end{equation}
and found rather similar behaviour to that seen for the GPEC action on
small (72 and 144 nodes) meshes - namely peaks in the specific heats
for the respective actions. For $S_1$ we have
\begin{equation}
C = {\lambda^2 \over N} \left( < S_{Steiner}^2 > - < S_{Steiner} >^2 \right).
\label{e11}
\end{equation}
We also see, by visual inspection of the surfaces, a low $\lambda$
crumpled phase and a large $\lambda$ smooth phase.
Although the gyration radius, a measure of the size
of the embedded surfaces,
\begin{equation}
X2 = { 1 \over 9 N (N -1)} \sum_{ij} \left( \vec X_i - \vec X_j \right)^2
q_i q_j
\label{e12}
\end{equation}
where the $q_i$ are the number of neighbours of point $i$,
was not monotone increasing with $\lambda$ as for the GPEC action
this could be explained by noting that the Steiner term, unlike
the extrinsic curvature, was dimensionful.

\section{Expected Scaling Properties}
Our simulations described above were carried out on boundaryless
surfaces with spherical topology. The simulations of \cite{2},
which were the most convincing demonstration to date of a
non-trivial continuum limit, required surfaces with boundaries
in order to carry out the scaling measurements of the
string tension and mass-gap. These are constructed in an
ingenious manner using twisted boundary
conditions on a torus, which we now outline, in order to avoid
pinning down a disproportionately large amount of lattice
at the boundary loops or points. It was observed in \cite{2}
that on a torus the sum of vectors $\vec X_{ij}$ along the edges of
the triangulation on a closed path could take the
values
\begin{equation}
\vec E( n_1 , n_2 ) = n_1 \vec E_1 + n_2 \vec E_2
\end{equation}
where the vectors $\vec E_1, \vec E_2$ are constant
and the integers $n_1, n_2$ denote how many times
the path winds round the two respective periods of
the torus. For non-zero $E_i$ this means that
\begin{equation}
\vec X_i ( k_1 , k_2 ) = \vec X_i + k_1 \vec E_1 + k_2 \vec E_2
\end{equation}
where the $k_i$ labelled the particular ``copy'' of the surface at a given
point. The partition function in equ.(2) is now dependent on
the choice of $\vec E_i$, $Z_N(\lambda) \rightarrow Z_N(\lambda, \vec E_1,
\vec E_2)$. Non-zero values of $\vec E_1, \vec E_2$ correspond
to simulating the surface on a frame $\vec E_1 \times \vec E_2$. The important
point to note is that it is not necessary to
designate any of the points as boundary points in this procedure.
It is thus possible to avoid potential poblems with too many points
on the boundary for small surfaces.

A canonical string tension
$\sigma (\lambda, N, y^2)$ for the system described above is
defined by taking $\vec E_1 = (y, 0, 0)$, $\vec E_2 = (0, y, 0)$,
$F_N(\lambda, y^2) = - \log Z_N(\lambda, y^2)$
and
\begin{equation}
\sigma (\lambda, N, y^2) = {\partial F_N(\lambda, y^2) \over
\partial y^2}
\end{equation}
where the translational invariance of $Z_N$ means it depends
on only the projected area $y^2$. Similarly a
canonical massgap is
defined by choosing $\vec E_1 = (y, 0, 0)$, $\vec E_2 = (0, 0, 0)$
and
\begin{equation}
m(\lambda, N, y) = {\partial F_N(\lambda, y) \over \partial y}.
\end{equation}
It is expected
that the $N$ and $y^2$ dependence in the string tension appears
as the ratio $r = y^2 / N$ and in the massgap as the ratio $t= y/N$.

It is actually more natural to define the physical string tension
in a grand canonical ensemble (with a varying number of points)
\cite{14},
which can be done by taking the Legendre transform of $F_N(\lambda,
y^2)$
\begin{equation}
G(\mu, \lambda, y^2) = N \mu  + F_N(\lambda, y^2)
\end{equation}
where $\mu$ is the cosmological constant. For large $y^2$
one expects $G(\mu, \lambda, y^2) \simeq \bar \sigma (\lambda, \mu) y^2$
with
\begin{equation}
\bar \sigma (\lambda, \mu) = { \partial F_N  \over \partial y^2} =
\sigma (\lambda, r).
\end{equation}
The grand canonical $\bar \sigma (\lambda, \mu)$ is expected
to scale as
\begin{equation}
\bar \sigma (\lambda, \mu) \simeq \sigma_0 (\lambda) + d ( \lambda)
\mu_R^{2 \nu}
\end{equation}
where the exponent $\nu$ governs the scaling of the physical
area $A_{phys} \simeq \mu_R^{2 \nu} y^2 $, with $\mu_R = (\mu -
\mu_{crit})$.
It is then possible to deduce the expected scaling of the {\it canonical}
$\sigma(\lambda, r)$:
\begin{equation}
\sigma(\lambda, r) \simeq \sigma_0 (\lambda) + \sigma_1 (\lambda) r^{2
\nu / ( 1 - 2 \nu) }.
\end{equation}
The physical string tension $\sigma_{phys} = \bar \sigma(\lambda, \mu) /
\mu_R^{2 \nu}$ will be infinite
unless $\sigma_0 (\lambda) \simeq (\lambda - \lambda_{crit})^{\alpha}$
as we approach a critical point at some $\lambda_{crit}$ and this
$\sigma_0$ is accessible in a canonical simulation.

It is also possible to play a similar game with the massgap, defining
\begin{equation}
G(\mu, \lambda, y) = N \mu  + F_N(\lambda, y)
\end{equation}
which is expected to behave as $G(\mu, \lambda, y) \simeq \bar m(\lambda, \mu)
y$ for large $y$. In this case we have $\bar m (\lambda, \mu) \simeq
\mu_R^\nu$ and
\begin{equation}
m(\lambda, t) \simeq D(\lambda) t^{\nu / (1 - \nu)}
\end{equation}
which is again accessible to a canonical simulation.

\section{Numerical Simulations and Results}
In this paper we apply the methods of \cite{2} to analyse the scaling of
the massgap and string tension for one of the variant Steiner
actions, $S_1$. This was chosen because the two terms
in the action ${1 \over 2} \sum_{<ij>} | X_i^{\mu} - X_j^{\mu} |$
and $\sum_{<ij>} | X_i^{\mu} - X_j^{\mu} | \theta (\alpha_{ij})$
have the same scaling dimensions which simplifies somewhat the
choice of observables. If we rescale the coordinates $\vec X_i \rightarrow y
\vec X'_i$ in $S_1$ we find $S_1(\vec X_i, y) \rightarrow y S_1 (\vec
X'_i , 1)$, which means that with
the appropriate boundary conditions for the string tension measurements
\begin{equation}
\sigma(\lambda, r) = { \partial F_N(\lambda, r) \over \partial y^2}  =
{<S_1> - 3 (N - 1) \over 2 y^2}
\end{equation}
for a surface with $N$ points embedded in 3 dimensions. If we use
the boundary conditions that are appropriate for the massgap
measurements we find
\begin{equation}
m(\lambda, r) = { \partial F_N(\lambda, t) \over \partial y}  =
{<S_1> - 3 (N - 1) \over  y}.
\end{equation}
We thus simply measure the expectation value of the
action with the appropriate choice of frame in order to access
information about the string tension and massgap scaling.

In addition, we measure the specific heat as defined in equ.(\ref{e11})
and histogram the output data at the various $\lambda$ simulated
in order to use the multi-histogram method of Ferrenberg and Swendsen
\cite{15}, which allows one
to estimate the density of states and hence the specific
heat for arbitrary $\lambda$. We also measure the gyration
radius, as defined in equ.(\ref{e12}), but a certain amount of
care is needed with this because of the twisted boundary conditions.
We choose to measure the $X2$ using only the component transverse
to the frame in the string tension measurements to avoid
confusion, and the two components transverse to the line
separating the pinned points in the case of the massgap measurements.
The autocorrelation times for the various observables
are calculated in order to ensure that we have
reasonable statistics.
We also measure the various acceptances for the lattice and $X$
moves to check that the Monte-Carlo algorithm, which we now describe,
is behaving reasonably.

In order to achieve a reasonable amount of vectorization in the code
64 systems were simulated in parallel, with measurements being taken
after a sufficient number of sweeps were made to allow
them to decorrelate. It proved to be convenient to
store the link variables $\vec X_{ij}$ rather than the site variables
$\vec X_i$ which allows the incorporation of
the twisted boundary conditions as
$\vec X_{ij} = \vec X_i - \vec X_j + \vec E_{ij}$, where
$\vec E_{ij} = n^1_{ij} \vec E_1 + n^2_{ij} \vec E_2$.
The integers $n_{ij}$ are non-zero when the link $<ij>$ passes
from one of the elementary cells in the parameter space
(a plane for the torus) to another.
Rounding errors during the simulation can be kept under control
by using the transformations
\begin{eqnarray}
\vec X_i &\rightarrow& \vec X_i + l^1_i \vec E_1 + l^2_i \vec E_2
\nonumber \\
\vec E_{ij} &\rightarrow& \vec E_{ij} + l^1_i \vec E_1 + l^2_i \vec E_2
- l^1_j \vec E_1 - l^2_j \vec E_2
\end{eqnarray}
where the $l$'s are arbitrary integers to keep the $E_{ij}$'s from straying.

The sum over lattices is effected by carrying out local flip moves
on adjacent triangles, forbidding flips that lead to
degenerate triangulations with less than 3 neighbours per point
or with 2-loops. With non-trivial boundary conditions the $\vec E_{ij}$
for affected edges must be changed, whereas the $\vec X_i$ are left
untouched. The coordinates $\vec X_i$ are updated with a simple
Metropolis scheme, which does not affect the non-trivial boundary
conditions. In this paper we report on simulations carried out
on relatively small surfaces of size 64 and 144 nodes. We
have not proceeded to larger surfaces in the current batch of
simulations because there is a hidden penalty built into the direct
transcription of the Steiner action we have used in $S_1$, compared
with the GPEC action. Namely, the calculation of $\theta ( \alpha_{ij} )$
requires an inverse trigonometric operation, rather than the simple
multiplications involved in calculating $\vec n \cdot \vec n$
in the GPEC action. It might be possible to avoid this
in further simulations by using
some trigonometric function with the requisite properties for
$\theta$ ($\theta( 2 \pi - \alpha) = \theta ( \alpha)$, $\theta (\pi )
= 0$, $\theta ( \alpha ) \ge 0$ ), but this begs the question
of universality.

If we now move on to discuss the measurements made for the
string tension and massgap with the choice of frames
$\vec E_1 = ( y , 0, 0 )$, $\vec E_2 = ( 0, y, 0)$ and
$\vec E_1 = (y, 0, 0)$, $\vec E_2 = ( 0, 0, 0 )$
respectively, comparison of Fig.1 for the string tension
and Fig.2 for the massgap with Figs.6,7 in the first of \cite{2} reveal
striking
similarities.
Looking at Fig.1 for the string tension
first we see that the data points, as expected,
fall on universal curves as a function of $r$ for a given $\lambda$
until finite size effects set in at small $r$.
Lines are drawn to guide the eye through the points coming from
the $N=64$ ($8 \times 8$) surface.
For large $r$, just as for the GPEC action, we would expect a $\lambda$
independent limit which is what is observed. For small $r$ the $\lambda$
dependence becomes more marked and as the $\lambda = 3, 4$ values straddle
zero as $r \rightarrow 0$ we can infer there is some $\lambda$ value
intermediate between these, say $\lambda_c$
where $\sigma (\lambda , r ) \rightarrow 0$
as $\lambda \rightarrow \lambda_c$ and $r \rightarrow 0$. This is
one of the prerequisites for a finite physical string tension.
For $\lambda = 4$ we see negative $\sigma ( \lambda , r )$ at small $r$,
which is due to the repulsion of the vertices, and the value where it is
zero corresponds to the equilibrium configuration. This is again similar
to the behaviour observed in the GPEC action, as are the very long
autocorrelation times observed in this phase.

We would expect the results for $m ( \lambda , t )$
in Fig.2 to fall on universal curves for different $\lambda$
with $t = y / N$ until finite size effects set in at small $t$, and
this is, indeed, what is seen. We have again drawn lines
through the $N=64$ ($8 \times 8$) points to guide the eye.
Only the crudest of fits to the scaling law in equ.(18)
($m \simeq D (\lambda ) t^{\beta (\lambda)}$, where $\beta(\lambda) =
\nu (\lambda) / ( 1
- \nu (\lambda))$ ) is possible with our data, and we find $\nu$ to be in the
region of $0.3$ for
$\lambda =3$ and $0.2$ for
$\lambda = 4$. This gives an estimate of $\nu$ at $\lambda_C$
which is lower than that for the GPEC action, $0.38< \nu < 0.42$.
This difference could be due to finite size effects.
It has recently been pointed out that these may be so strong
that they completely mask the presence of the tachyon,
and hence the real continuum physics,
for multiple spin models on dynamical lattices up to very large
sizes \cite{GH}, though the effects may well be less severe
here as they depend sensitively
on the extrinsic Hausdorff dimension (of the order
of 4 in this case) and probably non-universal constants.
We note
that our value is consistent with an earlier estimate for the GPEC action
in the second of \cite{2},
which was obtained using system sizes similar to those studied here.
A fit to the scaling law in equ.17 for $\sigma ( \lambda , r )$
gives a lower estimate for $\nu \le 0.2 $ for both $\lambda = 3, 4$
but the quality of the fits here is even poorer, so the discrepancy
is not too disconcerting.

The results for the specific heat are shown in Fig.3
for some selected $r$ values, where we have not
plotted the Ferrenberg-Swendsen interpolation between the measured
points for simplicity. The Ferrenberg-Swendsen interpolation
gives a value for $\lambda_C$ that is close to 4.
The quality of this interpolation deteriorates
as the relative size of the frame (ie $r$ or $t$) is increased,
suggesting that finite size effects may be quite important in such
cases. Finally the time evolution of $X2$ for $r=1$ is shown in
Fig.4 for one of the square
frames used in the string tension
measurements, and is at first sight rather surprising as it
{\it decreases} with time after a disordered start for large
$\lambda$. However, it should be remembered that  we measure
transverse fluctuations to the frame, so if the surface
becomes more rigid with increasing $\lambda$ this is what
we would expect to see as the transverse fluctuations are
suppressed.
When the frame size is reduced sufficiently the customary behaviour of
a larger $X2$ in the large $\lambda$ phase reasserts
itself.

To summarize the numerical results of this paper: for the
action $S_1$, containing an edge length and a Steiner term,
the scaling behaviour of both the string tension and the massgap
are very similar to those seen in the GPEC action. There is
some evidence for universal behaviour in
the value of the exponent $\nu$, but our measurements
of this are very preliminary. The behaviour of the specific heat peak is
consistent with that seen in our earlier, smaller scale, simulations of
$S_1$. Finally $X2$, both with and without framing the surfaces, gives
every indication of a smooth or rigid phase at larger $\lambda$.
There is no sign, however, of sharper scaling behaviour than is seen with
GPEC actions, for $S_1$ at any rate.
For future work it is possible that a
subdivision invariant action such as $S_2$ might offer a faster
approach to the continuum limit \cite{13}. A more judicious choice of
$\theta$ from the numerical point of view for any of the
Steiner actions $S_1, S_2, S_3$  might also offer the possibility of more
efficient
simulations.
Nonetheless, the current batch of simulations has demonstrated that there
is evidence for scaling and hence a non-trivial continuum theory for
a particular Steiner action, just as with the GPEC action.

\section{Acknowledgements}
The simulations were carried out on the Cray Y-MP at the
San Diego Supercomputer Centre, USA; the Cray Y-MP at the
Rutherford Lab, England (SERC grants GR/H 54904 and GR/J 21941);
and the Cray Y-MP at HLRZ
J\"ulich, Germany.
This work was supported in part by NATO collaborative research grant CRG910091
(CFB and DAJ),
and
by ARC grant 313-ARC-VI-92/37/scu (WJ and DAJ).
CFB was supported by DOE under contract DE-FG02-91ER40672
and by NSF Grand Challenge Applications
Group Grant ASC-9217394. WJ
thanks the Deutsche Forschungsgemeinschaft for a Heisenberg fellowship.
DAJ is supported at LPTHE
by an EEC HCM fellowship, an EEC HCM network grant and an Alliance grant.

\vfill
\eject

\vfill \eject
\centerline{\bf Figure Captions} \begin{description}
\item[Fig. 1.] The canonical string tension
$\sigma ( \lambda , r )$ is plotted for
the various $\lambda$ values to show the scaling with $r = y^2 / N$.
\item[Fig. 2.] The canonical massgap
$m ( \lambda , t )$ is plotted for the various
$\lambda$ to show the scaling with $t = y / N$.
\item[Fig. 3.] The specific heat vs $\lambda$ for some selected
$r$ values. Our original small scale simulations of $S_1$ \cite{10}
are also included for comparison.
\item[Fig. 4.] The {\it decrease} of the perpendicular $X2$
with time at large $\lambda$ is evident from these measurements taken
with $r = 1.0 $
\end{description}
\end{document}